\documentclass[prl,aps,floatfix,superscriptaddress,twocolumn]{revtex4}

\usepackage{amssymb,amsmath}
\usepackage{graphicx}

\begin{document}

\title{Turbulent black holes }

\author{Huan Yang,$^{1,2}$ Aaron Zimmerman,$^{3}$ Luis Lehner$^{1}$\\
%
$^{1}$Perimeter Institute for Theoretical Physics, Waterloo, ON N2L2Y5, Canada\\
$^{2}$Institute for Quantum Computing, University of Waterloo, Waterloo,
ON N2L3G1, Canada\\
$^{3}$ Canadian Institute for Theoretical Astrophysics, 60 St. George Street,
Toronto, On M5S3H8, Canada
}

\date{\today}

\begin{abstract}
We show that rapidly-spinning black holes can display turbulent
gravitational behavior which is mediated by a new type of parametric instability.
This instability transfers energy from higher 
temporal and azimuthal spatial frequencies to lower frequencies--- 
a phenomenon reminiscent of the inverse energy cascade displayed by $2+1$-dimensional turbulent
 fluids. Our finding reveals a path towards gravitational turbulence for 
rapidly-spinning black holes, and provides the first evidence for gravitational 
turbulence in an asymptotically flat spacetime. Interestingly, this finding
predicts observable gravitational wave signatures from such phenomena
in black hole binaries with high spins and gives a gravitational
description of turbulence relevant to the fluid-gravity duality. 
\end{abstract}

\maketitle

Black holes are fascinating objects. They play a fundamental role in a plethora of energetic phenomena in our universe, for example as the engines of active galactic nuclei, X-ray binaries, and possibly even as regulators of galactic structure. In addition, they have become central tools in the study of field theories through the framework of holography~\cite{Maldacena:1997re}.
This includes attempts to understand superfluidity, superconductivity and quark-gluon plasmas
obtained in energetic collisions 
(see e.g.~\cite{Chesler26072013,Horowitz:2010gk,CasalderreySolana:2011us}). One particularly exciting connection inspired
by holography is the ``fluid-gravity'' duality, which indicates
the dynamics of black holes in asymptotically anti-deSitter (AAdS) spacetimes in $d+1$
dimensions can be mapped
to the physics described by conformal fluids governed by viscous, relativistic 
hydrodynamics in $d$ dimensions~\cite{Policastro:2002se,Bhattacharyya:2008jc}. This opens the door to
search for particular behavior known to exist on one side of the duality on the other.
For instance, this duality has motivated studies showing that particular gravitational scenarios can
become turbulent when their fluid counterparts have high Reynolds 
numbers~\cite{Carrasco:2012nf,Adams:2013vsa,Green2014}. Additionally, concepts
in hydrydonamics, such as {\it ensthropy},
have geometric counterparts related to curvature quantities~\cite{Carrasco:2012nf}. This duality can also shed light
on poorly understood phenomena from a new perspective. Analyzing turbulence
from an intrinsically gravitational point of view is thus an exciting prospect.

In this work we develop a method to do precisely this and 
consider realistic, asymptotically flat black holes. Our analysis describes 
how gravitational turbulence
is mediated by a parametric instability in the gravitational field --which does not require
the ``confining properties'' of asymptotically AdS spacetimes-- and motivates the definition of a
{\em gravitational Reynolds number}. 
We first review general properties of turbulent flows, salient features of the fluid-gravity duality, and parametric instability.

\noindent {\bf Hydrodynamic turbulence.} Turbulence is a ubiquitous property of fluid 
flows with sufficiently high Reynolds 
number  $({\rm Re} \equiv \rho/\eta v \lambda >> 1)$~\cite{landau_fluid_1987,boffetta2012two}. 
Here $v$ and $\lambda$ refer to the typical velocity and wavelength of characteristic modes of the solution,
and $\rho, \eta$ the fluid density and viscosity.
At high ${\rm Re}$, nonlinear interactions prevail over dissipation due to viscosity, and chaotic behavior
ensues. 
Turbulence displays several features: (i) an energy cascade (which can be towards
higher frequencies in 3-spatial dimensions or lower ones in 2-spatial dimensions), (ii) an exponential growth
--possibly transitory-- of additional modes in the solution and (iii) a breaking
of initial symmetries of the flow, which are only recovered in a statistical sense at later times.
Further, in the absence of a driving force, 
global norms of the solution
display a transient power-law decay, and
viscous losses then decrease ${\rm Re}$ until turbulence ends.
Beyond these broad aspects, a full understanding of turbulence is missing. 
A promising new road of study
has been furnished
through the fluid-gravity duality, provided a purely gravitational model for turbulence is available.
Here we develop such a model and uncover possibly astrophysical consequences. 

\noindent {\bf Fluid-gravity duality and black holes in AAdS vs AF.}
The fluid-gravity duality indicates long-wavelength perturbations of 
black holes in AAdS spacetimes can be described by relativistic hydrodynamic equations (with
an equation of state given by $p=\rho/d$)~\cite{Bhattacharyya:2008jc}. 
In addition to connecting known hydrodynamic and gravitational effects, such as loss of energy through the black hole horizon to viscous dissipation, the duality can reveal new phenomena.
The presence of turbulence in hydrodynamics indicates that a similar behavoir appears in 
perturbed AAdS black holes, and
this expectation has been confirmed 
by simulations 
of the gravitational side of the problem~\cite{Adams:2013vsa} which are direct counterparts
of those in the hydrodynamical front~\cite{Carrasco:2012nf,Green2014}.
Nevertheless, an analytical understanding of what mediates turbulence in gravity is an open question, as well as whether
such striking behavior can take place in the realistic case of asymptotically 
flat (AF) spacetimes.

In considering these questions we recall the differences in how these two classes of spacetimes relate to hydrodynamics.
Regardless of the class considered, a gradient expansion of the Einstein equations for 
long-wavelength perturbations of black holes gives rise to relativistic hydrodynamic
equations on a timelike hypersurface
~\cite{1986bhmp.book.....T,2010JHEP...03..063E}. 
However, only AAdS has a unique surface, lying at infinity, where the correspondence can be defined unambiguously. 
In both classes, perturbed black holes have a spectrum of free, damped oscillation modes known 
as {\it quasinormal modes}~(QNMs, see e.g.~\cite{KokkotasSchmidt1999,Berti2009}). 
Black holes in AAdS only lose energy through the event horizon 
(as its boundary acts as a confining box), while energy in AF spacetimes  can
also be lost to infinity. Consequently, QNMs decay considerably more
slowly in the AAdS case. From the hydrodynamic view, a slow decay of QNMs
implies low viscosity and a correspondingly higher Reynolds number~\cite{Green2014}.
In what follows, we show that this slow decay is key for generating turbulent behavior, and 
how it might arise in the AF case. By doing so, we provide the first gravitational description of a turbulent mechanism acting in realistic black hole spacetimes. 

\noindent{\bf Damped parametric oscillator.} 
The parametric instability in black holes described below is analogous to the simple parametric oscillator. 
A parametrically driven oscillator can be described by the equation
\begin{equation}
\ddot q +\gamma \dot q+\omega^2\left [1+ 2 f(t) \right ]q=0\,,
\end{equation}
where $\omega$ is the intrinsic harmonic frequency, $\gamma$ is a weak damping coefficient ($\gamma \ll \omega$) and $f(t)$ characterizes the parametric driving. The solution to this equation is bounded in time, except when $f(t)$ oscillates at approximately twice the intrinsic frequency: $f(t)=f_0\cos \omega' t,\,\omega' \approx 2 \omega$. In this case the time dependence of the solution is described 
by $e^{\Omega t}$, with the rate $2 \Omega \approx \omega \sqrt{f_0^2-\omega^{-4} [\omega^2-(\omega'/2)^2]^2}-\gamma .
$
When $\omega'$ is close to $2\omega$, a small parametric driving amplitude $f_0$ will be able to excite a growing solution, which is referred as a {\it parametric instability}. For a given 
value of the damping coefficient $\gamma$, there is a critical relation 
that $f_0$ and $\omega$ satisfy at the separatrix between
growth or decay. This is related to the critical gravitational Reynolds number for the onset 
of turbulent behavior in perturbed black holes. 

\noindent {\bf Perturbed black holes in AF scenarios and turbulence.}
In 4 dimensions, a stationary AF black hole is characterized by its mass $M$ and spin parameter $a$, 
which has a maxiumum value of $a/M  = 1$. When $a/M \approx 1$ or $\epsilon \equiv 1-a/M \ll 1$,
 there exists a family of quasinormal modes with a small damping rate proportional to
 $\sqrt{\epsilon}$ (referred as zero-damping-modes or ZDMs)~\cite{Detweiler1977,Hod2008a,Yang:2012pj,Yang:2013uba}. These modes have time dependence $e^{i \omega_{lmn} t}$, with
\begin{equation}
\label{eq:ZDMfreq}
\omega_{lmn}\equiv\omega_R-i \omega_I\approx \frac{m}{2}-\frac{\delta \sqrt{\epsilon}}{\sqrt{2}}-i\left (n+\frac{1}{2}\right ) \frac{\sqrt{\epsilon}}{\sqrt{2}} \,,
\end{equation}
(with $l,m,n$ denoting the angular, azimuthal and overtone numbers respectively, and $\delta$
a function of $l,m$ and the spin-weight of the perturbation considered, see Supplemental Material).
Consider as an example a black hole perturbed by a small mass falling towards 
the event horizon. This excites some of the ZDMs to a characteristic amplitude $h_0$. 
Once a particular ZDM is excited, at linear order its amplitude
decays exponentially with a rate $\propto \sqrt{\epsilon}$ (in hydrodynamical terms this decay corresponds to laminar flow). However, nonlinear coupling between modes introduces a competing energy transfer between modes at a rate dependent on $h_0$.
As we decrease $\epsilon$, the mode-mode coupling mechanism may overcome decay,
even pumping up modes that are not initially excited,
regardless of how weak the initial perturbation is. This is analogous to the onset of 
turbulence at high ${\rm Re}$. 

\noindent {\bf Formalism.} As we go beyond linear perturbation theory, the spacetime metric $g$ can be 
expanded as $g=g_B+h^{(1)}+h^{(2)}+...\,$, where $g_B$ is the background Kerr metric 
and $h^{(n)}$  is the $n$th order perturbation.
We are interested in how an initial ZDM metric perturbation $h^{(1)}$ might trigger
other modes through parametric resonance. 
One way to analyze the problem is to take $g_{\tilde B}=g_B+h^{(1)}$ as a dynamical background metric
 and study the evolution of $h^{(2)}$ on it. To avoid delicate gauge issues
 for the higher order metric perturbations, we adopt a simpler version of this approach,  solving
the evolution of a massless scalar field in the dynamical background $g_{\tilde B}$. This field obeys the wave equation
\begin{align}
\Box_{\tilde B} \Phi=0\,,
\end{align}
and we bear in mind that $\Phi$ plays a role analogous to $h^{(2)}$. 
Since $\Box_{\tilde B} \Phi$ is gauge invariant, our results concerning the parametric instability are gauge invariant.

The first-order perturbation $h^{(1)}$ corresponding to a quasinormal mode with index $(l,m,n)$ is
\begin{align}\label{eqboxexp}
h^{(1)}_{\mu\nu} = 2 h_0\,{\Re}\left [Z_{\mu\nu}(r,\theta)e^{-i\omega t+i m \phi}\right ]  \,, 
\end{align}
where $h_0(t)=h_0e^{-\omega_I t}$.
As we perturb the background metric $g_B$ to $g_B+h^{(1)}$, $\Phi$ obeys,
\begin{align}\label{eqboxop}
\Box_{\tilde B} \Phi 
\approx \left [ \Box_B+\frac{1}{\Sigma} \mathcal{H}(h^{(1)}) \right ] \Phi.
\end{align} 
Here $\Sigma \equiv r^2+a^2\cos^2\theta$ and $\mathcal{H}(.)$ is a time-dependent operator linear in its argument. The time dependence of $\mathcal H$ is crucial in triggering the parametric instability, which occurs when the 
temporal and azimuthal frequencies of the parent $h^{(1)}$ match the daughter mode $\Phi$. For rapidly-spinning Kerr black holes, this occurs when the 
the daughter mode
satisfies $m'=m/2$, as Eq.~\eqref{eq:ZDMfreq} guarantees that $\omega'_R \approx \omega_R/2$ as well. 
We make the ansatz
$$
 \Phi_{l'm'n'}(x^\mu) = \left [g_j(t)e^{(-1)^j i\omega_R/2 t - (-1)^j i m' \phi}Y_{l'm'n'} \right ]e^{- \omega'_I t} ,\\
$$
(summing over $j=1,2$), with ${g_1,g_2}$ characterizing the time dependence and $Y_{l'm'n'}(r,\theta)$ the perturbed
wave function. The equations of motion determining
$g_1,g_2$ are closely related to the parametric instability previously discussed. 
The solution to these equations are given by $g_j = A_j e^{\int \alpha(t') dt'}$ with
\begin{align}
\label{eq:alpha}
\alpha = \pm\sqrt{\left |{H h_0(t)}/{Q m'} \right |^2-\left (\omega'_R-{\omega_R}/{2} \right )^2}\,,
\end{align}
where $H$ has the physical meaning of mode-mode coupling strength and $Q$ gives the susceptibility of the wave equation to a perturbation of the mode frequency. At leading order, $Q$ is independent of $m$.
An exponential growth in $\Phi$ will occur if $\Omega \equiv \alpha(t) - \omega'_I > 0 $, i.e. when
\begin{align}
\label{eqcri}
 {h_0(t) }/(m' \omega'_I)  - \left |{Q }/{H } \right | \sqrt{\left (\omega'_R-{\omega_R}/{2} \right )^2/{\omega'}^2_I+1} \, > 0 \,. 
 \end{align}  
We emphasize that given $m'=m/2$, both $\omega'_R-\omega_R/2$ and $\omega'_I$ can be read off from Eq.~\eqref{eq:ZDMfreq}, and both are $\propto \sqrt{\epsilon}$. 
We choose to normalize the radial wave function of the ZDMs such that $|H/Q|$ is $\epsilon$ independent --- in other words, 
the effect of mode-mode coupling stays constant for varying black hole spins~\footnote{This normalization means that the amplitude $h_0$ of a particular mode excited by a physical process will have an additional dependence on $\epsilon$, see the Supplemental Material.}. These properties are useful in defining and interpreting the gravitational Reynolds number.  

\noindent{\bf Turbulent Black Holes.}
Based on the above analysis, consider an initial ZDM mode with $m=2m'$ and amplitude $h_0$; as we increase $h_0$, all the secondary ZDMs with azimuthal quantum number $m'$ satisfying Eq.~\eqref{eqcri} are parametrically excited. As these daughter modes 
grow, 
energy flows from the parent mode to the daughter modes, and the 
parent mode experiences back reaction due to the mode coupling.
Ignoring this back reaction, these secondary modes grow as long as Eq.~\eqref{eqcri} holds, but in a realistic situation
parametric growth terminates when the amplitudes of the parent mode and the secondary modes become comparable requiring a fully nonlinear treatment (or numerical study, e.g. \cite{East:2013mfa}).
The gravitational parametric instability displays an {\it inverse cascade}, as energy flows 
from modes with high azimuthal frequencies to modes with lower azimuthal 
frequencies, {\em and from higher to lower temporal frequencies}. An initial azimuthal mode $m$ generates a series of of modes with azimuthal number $m/2^p$ after $p$ generations. This is  
similar to the inverse energy cascade in $2+1$-dimensional turbulent fluids. 
Since modes with the same $m'$ but high $l$ can also be excited, there is also a direct transfer 
of energy towards higher overall angular frequencies. 

From the criteria in Eq.~\eqref{eqcri} we define a {\it gravitational Reynolds number} ${\rm Re}_g$, 
taking $m=2m'$, and with $\omega'_I$ chosen to be the lowest possible decay rate of all the 
ZDMs, $\gamma_\eta =\sqrt{\epsilon/8}$. This gives
\begin{align}
{\rm Re}_g \equiv h_0/(m\,\gamma_\eta). 
\end{align}
For a mode having ${\rm Re}_g$ below some critical value given in Eq.~(\ref{eqcri}), no growth is expected, and the mode so the mode behaves in a ``laminar'' manner, decaying normally.
For larger values of ${\rm Re}_g$, turbulent behavior ensues, driving growing modes and a richer angular structure. Once ${\rm Re}_g$ decreases below the critical value for a given mode, that mode again decays exponentially.
Notice that the natural identifications
$\{ \eta/\rho \leftrightarrow \gamma_\eta,\, L \leftrightarrow 1/m,\, v \leftrightarrow \,h_0\}$ gives 
$
{\rm Re}_g \leftrightarrow {\rm Re}.
$ 
Our definition arises from the criteria for the onset of instability, and it agrees with the one proposed in~\cite{Green2014} motivated
through the fluid-gravity duality. Table 1 presents a list of numerical values 
of the {\it critical} ${\rm Re}_g$, beyond which the parametric instability for different driving and secondary modes will be turned on. We consider only the lowest overtone modes, $n= n' =0$. We can see that for fixed
$\epsilon$ and $m$, the critical ${\rm Re}_g$ asymptotes to a constant value at for high $l$ modes. 
One may argue that this means modes with arbitrarily high $l$ are all excited. 
However, as discussed in Yang {\it et al.}~\cite{Yang:2012pj,Yang:2013uba} there is a minimum, 
critical $\epsilon$, beyond which the required phase-matching condition gradually fails to hold. A conservative estimate for this critical value 
is $\epsilon_c \propto l^{-2}$. So for a given spin, there is a high angular frequency cut-off
 scale where the instability criteria is not satisfied and the energy-transfer stops.

\begin{table}
\begin{tabular}{|c|c|c|c|c|c|c|c|c|}
\hline
$(l,m)$  & $l'=1$ & $l'=2$ & $l'=3$ & $l'=4$ & $l'=5$  & $l'=6$ & $l'=7$ &$l'=8$  \\
\hline
$(2, 2)$ & 0.287 & 0.163 & 0.130 & 0.122 & 0.117 & 0.115 & 0.113 & 0.111  \\
\hline
$(4, 2)$ & 43.2 & 62.1 & 92.7 & 123 & 118 & 118 & 117 & 117 \\
\hline
$(4, 4)$ & -- & 3.62 & 0.00676 & 0.0114 & 0.0108 & 0.0104 & 0.0101 & 0.0100  \\
\hline
\end{tabular} 
\caption{Critical ${\rm Re}_g$ for different parent daughter modes with $m=2m'$. These numbers are obtained in the ingoing radiation gauge using a value for $|H/Q|$ evaluated at $\epsilon = 10^{-5}$ (although they are expected to be $\epsilon$-independent, numerically we use a small $\epsilon$ to reduce systematic error in the wave functions); extrapolation to lower spins and error in the matching of radial eigenfunctions are the dominant sources of error, which we estimate conservatively to be 10\%. The parent mode of the $42 \to l1$ driving has an imaginary value of $\delta$, whereas the parents in the other two cases have real $\delta$, which may explain the large critical Reynolds numbers in those cases. Note also that the $44 \to 22$ driving is unique in the sense that both its parent and daughter mode have real $\delta$.}
\end{table}

Figure~1 illustrates the rich angular structure of the perturbed spacetime that arises due to the parametric instability, due to driving by the fundamental $l=2$, $m=2$, $n=0$ QNM. We take for our fiducial example $\epsilon = 2 \times10^{-3}$ ($a/M=0.998$)
\footnote{Note that our perturbative analysis is about an isolated black hole, and so $a$ corresponds to the spin parameter of the final black hole in the case of a binary merger.}
and $h_0(t=0) = (1/8) \sqrt{\epsilon}$. This amplitude is motivated by the expected excitation following a large mass-ratio inspiral, such as can occur in supermassive binary black hole
coalescence following galaxy mergers (see the Supplemental Material). Note that for such an $h_0$ the criteria for growth is independent of spin, so long as $\epsilon \ll1$.
In the fully gravitational case, we can expect a similar development of structure in both the far-field radiation and curvature quantities on the event horizon. 
Figure~2 shows the amplitudes of the driving gravitational mode and excited scalar modes for the same fiducial example as in Fig.~1. 
Though we focus on driving by the dominant $(2,2)$ mode, Table~1 indicates that modes can be driven by a $(4,4)$ mode for even smaller values of $h_0$.
\begin{figure}
\includegraphics[width=0.75\columnwidth]{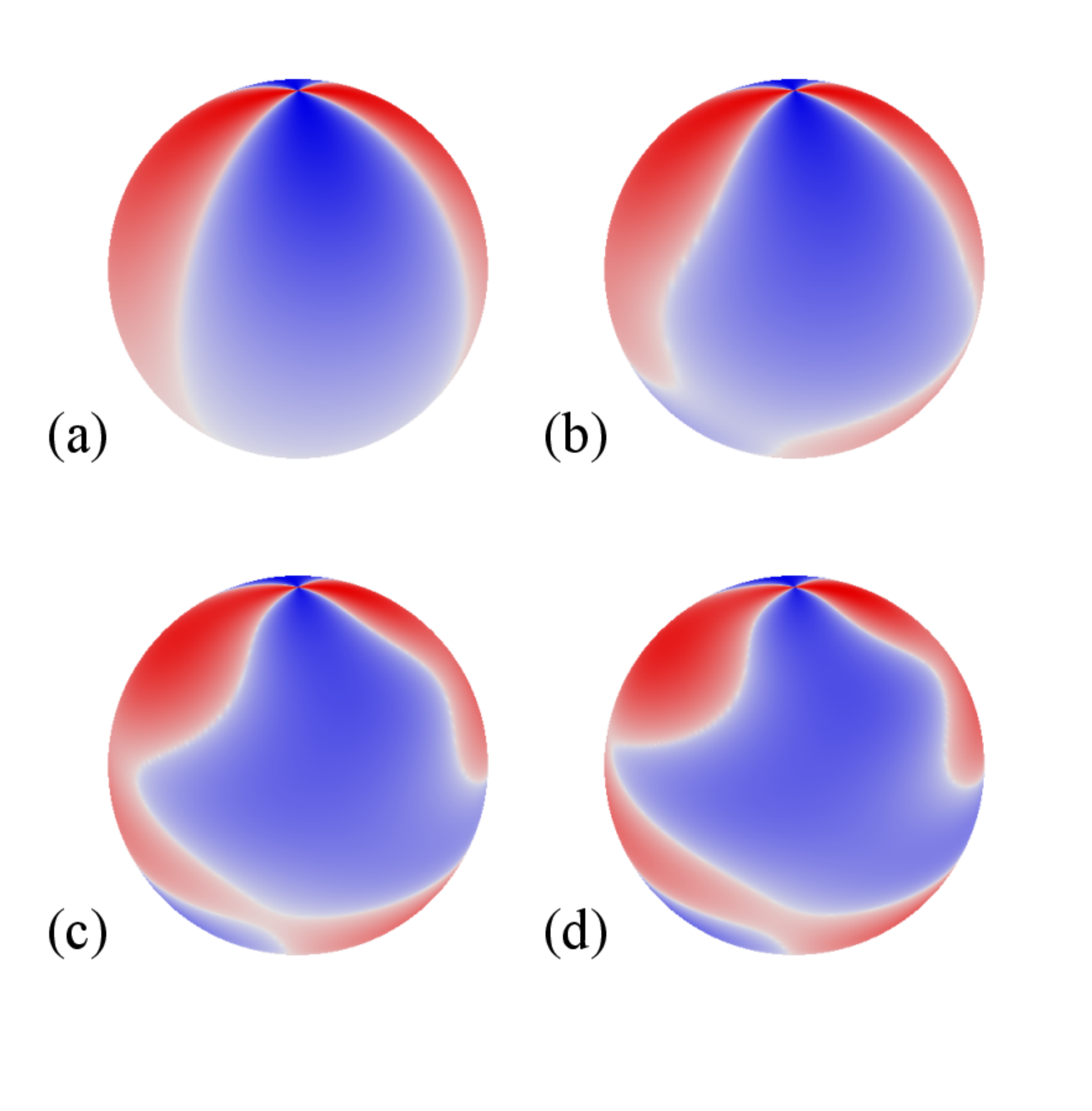}
\vspace{-6mm}
\label{fig:Spheres}
\caption{Snapshots of parametrically driven modes on a sphere of constant radius. We plot the dominant $(2,2)$ spin $s=-2$ spheroidal harmonic from a collective driving mode, plus the spin-0 $(l,1)$ spheroidal harmonics for all of the growing scalar modes.
Initially $h_0(t=0)= (1/8) \sqrt{\epsilon}$, and $\epsilon = 2 \times10^{-3}$ (a = 0.998). In this case, modes with $2\ \leq l \leq 6$ are resonantly excited, with the higher $l$ modes growing faster; the $l>6$ modes are not ZDMs for this $a$. At $t/M = 0$, the scalar modes are seeded with equal amplitude 10\% of the gravitational mode, and random phases.  
{(a)}  Reference spin $s=-2$, $(2,2)$ spheroidal harmonic. {(b)} At time $t/M = 0$, the seed modes are visible only where the gravitational mode is weak. {(c)} At time $t/M = 16$, more angular structure has developed. {(d)} The harmonics at $t/M=32$ when the amplitude of the $(6,1)$ scalar mode is closest to the $(2,2)$ mode.}
\end{figure}

\begin{figure}
\includegraphics[width=1.0\columnwidth]{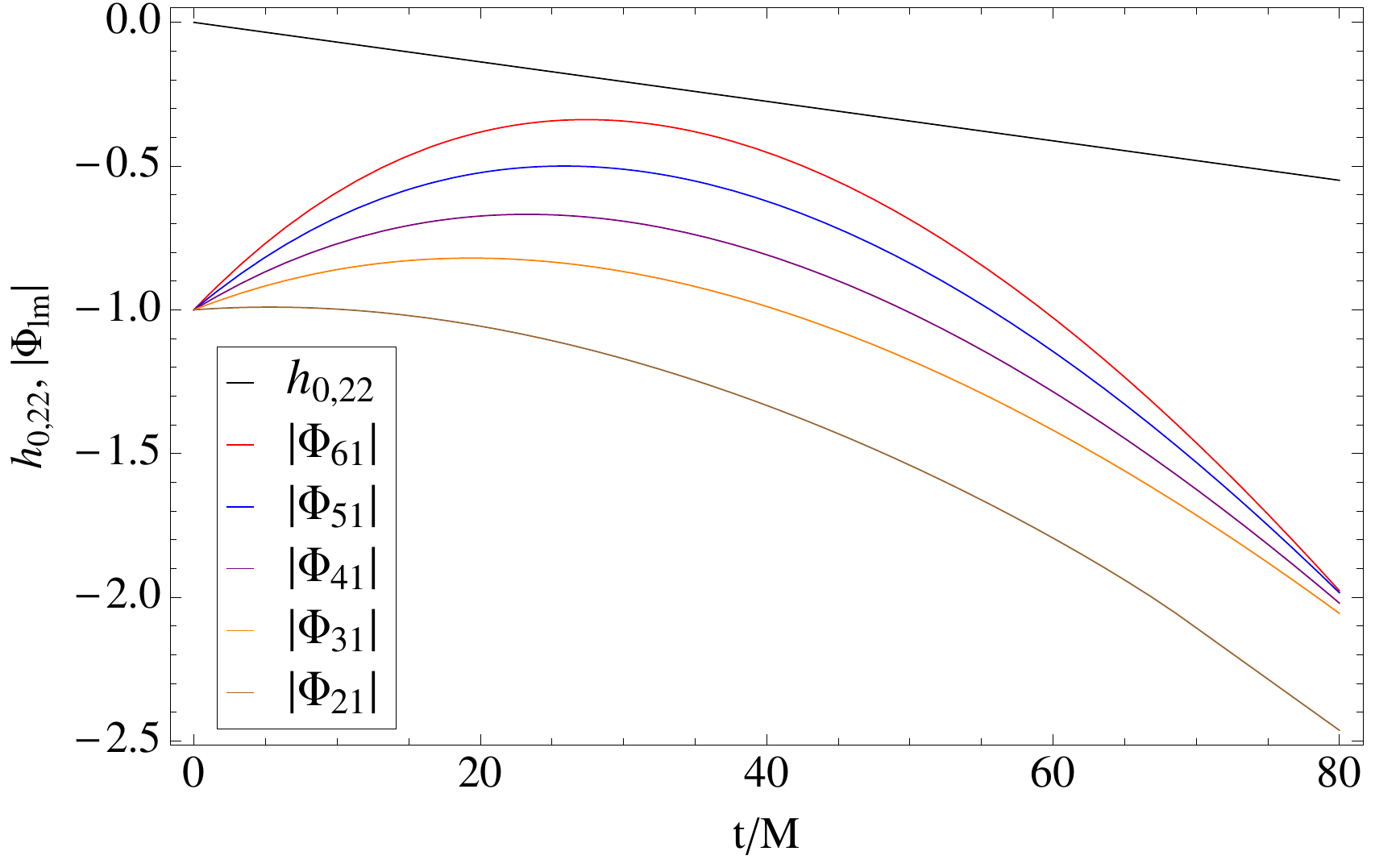}
\label{fig:gw}
\caption{Growth of the scalar quasinormal mode amplitudes due to a $(2,2)$ perturbation, on a logarithmic scale, using the same parameters as in Fig.~1. After initial parametric growth, the driving of each mode turns off as $h_0(t)$ decays, after which the scalar mode decays at its standard exponential rate, which is larger for modes with larger $l$.}
\end{figure}

During the inverse cascade, modes with frequencies $2^{-p}$ ($p \in \mathbb Z$) times the parent 
mode frequency are excited by parametric resonance. However, in a fully turbulent fluid, energy transfers 
throughout the entire spectrum. One possible mechanism for this is in the gravitational case is through {\it resonant excitation} of additional modes, as occurs in systems of coupled oscillators. For example, two oscillators with
frequencies $\omega_1$ and $\omega_2$, and amplitudes $A_1(t)$ and $A_2(t)$ can drive modes with frequencies $\omega' = \omega_1 \pm \omega_2$, resulting in amplitudes proportional to $A_1 A_2$. 
These three-mode interactions are not as strong as the parametric resonance but they can redistribute 
energy to both higher and lower frequencies, and fill in the gaps in the spectrum.

\noindent{\bf Observational consequences.}
This parametric instability discussed here relies on
the system having a rapidly spinning black hole. Theoretical models arguing for such scenarios
have been developed~\cite{1974ApJ...191..507T,Kesden:2009ds} and, crucially, there is observational evidence for highly spinning black holes~\cite{McClintock:2011zq,2013Natur.494..449R}. The turbulent instability
has several possible signatures: 
$\bullet$ {\bf Gravitational wave structure}. In gravitational wave observations 
from large mass-ratio mergers involving a rapidly-spinning black hole.
Such scenarios can arise for instance in the inspiral of supermassive binary black
holes following galaxy mergers.
After merger 
the final black hole rings down by emitting gravitational 
waves primarily through the $(2,2)$ mode. The magnitude of the initial 
perturbation $h_0$ is proportional to the mass ratio, and so for smaller $\mu$ values
 Eq.~\eqref{eqcri} is not satisfied, and distant observers should see mainly the 
$(2,2)$ mode during the entire ringdown. However, if the initial perturbation is
 strong enough, modes with $m=1$ will be parametrically excited. 
The growth 
of the modes can allow them to overtake the amplitude of the $(2,2)$ mode, in 
which case a treatment of the back reaction is needed. However, it is possible that a 
distant observer could measure a growing amplitude of some modes during the 
ringdown, a clear evidence of the instability, perhaps followed by
 complicated and turbulent behavior in the mode structure of the observed signal. 
Gravitational wave signals
from supermassive binary black hole mergers would be detectable by
pulsar timing arrays (e.g.~\cite{2013CQGra..30v4010I}) while
stellar mass systems are the target of 
LIGO/VIRGO/KAGRA~\cite{Abbott:2007kv,2011CQGra..28k4002A,Somiya:2011np}.
$\bullet$ {\bf Jitter in the black hole geometry.} The phenomena discussed indicates that
the geometry of the spacetime around a black hole can acquire a rich multipolar structure as
as a result of an object falling into a rapidly spinning black hole. This structure will impact
the surrounding region and, in particular, may cause angular time-dependent shifts in the location
of the inner most stable circular orbit. This, in turn, can affect emission lines of accreting material.
$\bullet$ {\bf Chaos in black holes?} 
We have seen that turbulent behavior occurs in nearly extremal black holes, where the mode-mode coupling concentrates 
near the horizon. This may be related to the fact that the black hole singularity moves ``closer" to the 
horizon for higher black hole spins, and the chaotic region near the 
singularity may be reflected in the existence of turbulence near the horizon.
Recently it has been suggested that chaotic 
behavior in the vicinity of the black hole singularity may be responsible for the information loss in 
the {\it black hole information paradox}~\cite{Hawking:2014tga}. Our work indicates that complicated behavior arises in a transitory way outside of the event horizon if the gravitational Reynolds number is high enough.

{\bf Acknowledgments.}
We thank Stephen Green and Scott Hughes for discussions, and Zachary Mark for his help 
in validating the inner product used here in a separate study. We also thank Eric Poisson for valuable comments on this manuscript. We thank Chris Thompson for discussion which led us to identify an error in an earlier preprint.
This work was supported by NSERC through Discovery Grants and CIFAR
(to L.~L.). This research was supported in
part by Perimeter Institute for Theoretical Physics. Research at
Perimeter Institute is supported by the Government of Canada through
Industry Canada and by the Province of Ontario through the Ministry
of Research and Innovation.

\section{Supplemental Material}

{\bf Perturbative formalism and metric reconstruction.} 
We are interested in the perturbations of a Kerr black hole beyond linear order, with the spacetime metric is expanded as $g=g_B+h^{(1)}+h^{(2)}+...\,$, where $g_B$ is the background metric for a hole of mass $M$ and spin parameter $a$, given by
\begin{subequations}
\label{eq:BL}
\begin{align}
ds^2  = & -\left( 1 - \frac{2 M r}{\Sigma} \right) dt^2 - \frac{4 M r a \sin^2 \theta}{\Sigma} dt d \phi + \frac{\Sigma}{\Delta} dr^2  \notag \\
&+ \Sigma  d \theta^2  + \left( r^2 + a ^2 + \frac{2 M r a^2 \sin^2 \theta}{\Sigma} \right) \sin^2 \theta d\phi^2 , \\
\Sigma =&  r^2 + a^2 \cos^2 \theta\,, \qquad \Delta =  r^2 -2 M r +a^2 ,
\end{align}
\end{subequations}
and $h^{(n)}$ is the $n$th order perturbation field with amplitude $\propto h^n_0$. 
Given the initial excitation of a ZDM with metric perturbation $h^{(1)}$, we wish to understand how other modes evolve when we take into account mode-mode coupling.
As a model for the problem of nonlinear mode coupling of the gravitational perturbations of a Kerr background, we consider the scalar wave equation
$\Box_{\tilde B} \Phi=0\,, $
in the dynamical background metric $g_{\tilde B}=g_B+h^{(1)}$. In this model, $\Phi$ is analogous to a higher order metric perturbation $h^{(2)}$, and we expect it to have the same qualitative behavior as the problem of interest. Under small gauge transformations $x^\mu \to x^\mu + \xi^\mu$, $\Phi$ has the simple transformation $\Phi(x) \to \Phi(x) + \xi^\mu \partial_\mu \Phi(x)$ to the order we are concerned with. We have adopted geometric units with $G=c=1$, and from here, we measure length in units of the black hole mass $M$, setting $M=1$.

The first-order perturbed metric $h^{(1)}$ corresponding to a quasinormal mode can be obtained from the Weyl scalar $\Psi_4$ (or $\Psi_0$) using a specific gauge choice. 
For a mode $(l,m,n)$ with amplitude $h_0$, $\Psi_4$ is given by~\cite{Teukolsky1973}
 \begin{equation}\label{eqpsi4}
 \Psi_4 =h_0 \,e^{- i \omega_{lmn} t + i m \phi} {}_{-2}S_{lmn}(\theta) {}_{-2}R_{lmn}(r)\,.
\end{equation} 
Here ${}_{-2}S_{lmn}(\theta)$ is spin-weighted spheroidal harmonic function (with spin weight $s=-2$) and ${}_{-2}R_{lmn}(r)$ is the radial wave function of the quasinormal mode.  The ZDM mode frequency is approximately~\cite{Hod2008a,Yang:2012pj,Yang:2013uba}
\begin{equation}
\label{eqzdmfreq}
\omega_{lmn}\equiv\omega_R-i \omega_I\approx \frac{m}{2}-\frac{\delta \sqrt{\epsilon}}{\sqrt{2}}-i\left (n+\frac{1}{2}\right ) \frac{\sqrt{\epsilon}}{\sqrt{2}} \,,
\end{equation}
where $\epsilon=1-a$, $\delta^2 \equiv 7m^2/4-(s+1/2)^2-A_{lm}$ and $A_{lm}$ is the eigenvalue of the spin-weighted spheroidal harmonic function. 

We construct the spin-weighted spheroidal harmonics at leading order in small $\epsilon$ using the power series expansion discussed by Leaver in~\cite{Leaver1985}. In this limit they are real. For the radial wave functions, we must use expressions for $R_{lm\omega}$ that are appropriate for nearly extremal Kerr black holes. We discuss them, their normalization, and the expected size of $h_0$ below, where we detail our method of constructing an appropriate inner product on the radial functions. 

With knowledge of $\Psi_4$, the corresponding metric perturbation $h^{(1)}$ can be reconstructed. In the Kerr spacetime, metric reconstruction is performed in one of two gauges --- the ingoing and outgoing radiation gauges, as first carried out by Chrzanowski~\cite{Chrzanowski:1975wv} and developed by others (see e.g.~\cite{Wald:1978vm,Lousto:2002em,Ori:2002uv,Keidl:2006wk,Keidl:2010pm,Nichols:2012jn}). The details of the metric reconstruction procedure are relatively lengthy and tedious, requiring the application of Newman-Penrose formalism~\cite{NewmanPenrose1962}, and so we will only summarize the major steps here. We compute $h^{(1)}$ in ingoing radiation gauge, using the standard Kinnersley null tetrad vectors $l^\mu, n^\mu, m^\mu, m^{*\mu}$, as discussed in~\cite{Nichols:2012jn}. The metric $h^{(1)}$ is built by applying a tensor differential operator to a scalar $\Psi_H$ known as the Hertz potential,
\begin{align}
h_{\mu \nu} = & \bigl(- l_\mu l_\nu ({\boldsymbol \delta} +\alpha^* + 3\beta - \tau)({\boldsymbol \delta} + 4 \beta + 3 \tau)  
\notag \\ &
 - m_\mu m_\nu ( {\bf D} - \rho + 3 \epsilon - \epsilon^*)({\bf D} + 3 \rho + 3 \epsilon) \notag \\
&  +l_{(\mu}m_{\nu)} \left[({\bf D} + \rho^* - \rho + \epsilon^* + 3 \epsilon)({\boldsymbol \delta} + 4 \beta + 3 \tau) \right.
\notag \\ & \left.
 + ({\boldsymbol \delta} + 3 \beta - \alpha^* - \pi^* - \tau)({\bf D} + 3 \rho +4 \epsilon) \right] \bigr) \Psi_H + {\rm c.c.} 
\end{align}
Here ${\boldsymbol \delta} = m^\mu \partial_\mu$, and ${\bf D} = l^\mu \partial_\mu$ are directional derivatives; $\alpha, \beta, \tau, \rho, \epsilon, \pi$ are the scalar Newman-Penrose spin coefficients for the Kerr spacetime and Kinnersley tetrad and can be found in e.g.~\cite{Teukolsky1973}; and c.c. indicates the complex conjugate of the preceding expression.
The Hertz potential which generates the desired $\Psi_4$ in ingoing radiation gauge is given by
\begin{align}
\Psi^{\rm IRG}_H  = \sum_{l m \omega} e^{-i \omega t}e^{i m \phi} {}_{-2} S_{lm \omega} (\theta) {}_{-2} X_{lm\omega}(r) \,, 
\end{align}
with a the radial function given by
\begin{align}
{}_{-2} X_{l m \omega} = 8 \frac{(-1)^m D_{lm\omega}^* - 12 i M \omega} {D_{lm\omega}^{*2} + 144 M^2 \omega^2} {}_{-2} R_{l m \omega} \,.
\end{align}
Here,
\begin{align}
D_{lm\omega}^2 =& \lambda_C^2 ( \lambda_C + 2)^2 - 8 \lambda_C(5\lambda_C + 6) (a^2 \omega^2 - am \omega) \notag \\ &
+ 96 \lambda_C a^2 \omega^2 + 144(a^2 \omega^2 - am \omega)^2 \,, 
\end{align}
and $\lambda_C = A_{lm} + s + |s| - 2 a m \omega +a^2 \omega^2$ is the angular separation constant used by Chandrasekhar~\cite{ChandraBook}, which differs from that originally used by Teukolsky~\cite{Teukolsky1973}. In deriving our simple expression for the radial part of $\Psi_H$, we have required that our Teukolsky radial function obeys the identity $R^*_{l -m - \omega^*} = R_{lm\omega}$, and is accomplished in our case by choosing the convention that for the $-m, -\omega^*$ mode, $\delta <0$ if $\delta$ is real and $\delta = i \delta'$ with $\delta'<0$ is $\delta$ is imaginary, whereas usually $\delta>0$ for real $\delta$ or $\delta'>0$ for imaginary $\delta$. \\

{\bf Parametric wave instability} --- Let us assume that initially a quasinormal mode with $\Psi_4$ given by Eq.~\eqref{eqpsi4} is injected into the black hole spacetime.
Because $\partial _t$ and $\partial_\phi$ are the two Killing-fields of the Kerr spacetime, after the metric reconstruction $h^{(1)}$ shares the same periodic $t$ and $\phi$ dependence as $\Psi_4$. We denote it as  \\
\begin{align}\label{eqboxexp}
 h^{(1)}_{\mu\nu} 
 = 2 h_0(t)\,{\Re}\left [Z_{\mu\nu}(r,\theta)e^{i(-\omega t+m \phi)}\right ] \,.
\end{align}
The time dependence of the amplitude is that of a single quasinormal mode, as discussed above.
As we perturb the background metric $g_B$ to $g_B+h^{(1)}$, the d'Alembertian in the new background becomes
\begin{align}\label{eqboxop}
\Box_{\tilde B} \Phi 
\approx & \Box_B \Phi- \frac{1}{\sqrt{-g_B}} \partial_\mu \left( h^{(1)\mu \nu} \sqrt{-g_B} \partial_\nu \Phi \right) \notag \\ &
+ \frac12 g_B^{\mu \nu} \left(\partial_\mu h^{(1)\rho}{}_\rho \right)  \partial_\nu \Phi\, 
\notag \\&
\equiv \left [ \Box_B+\frac{1}{\Sigma} \mathcal{H}(h^{(1)}) \right ] \Phi.
\end{align} 
Here $\mathcal{H}(.)$ is an operator linear in its argument. In the ingoing and outgoing radiation gauges, the metric perturbation is traceless, $h^\rho{}_\rho=0$, and so the $\mathcal{H}$ operator in 
Eq.~\eqref{eqboxop} simplifies. 

In analogy with the parametric oscillator, a parametric instability may occur if the driving frequency 
is approximately twice the intrinsic harmonic frequency. 
This condition can be generalized when studying parametric wave generation, in which case the temporal and azimuthal frequencies have to be simultaneously matched between
 the parent mode $h^{(1)}$ and the secondary mode $\Phi$. This requirement is difficult to meet for
 perturbations in generic Kerr black holes, but for the rapidly-spinning ones, if the secondary mode we consider
 satisfies $m'=m/2$, then Eq.~\eqref{eqzdmfreq} will guarantee that $\omega'_R \approx \omega_R/2$ as well. 
In other words, a generalized matching condition holds for $\Phi$ modes $(l', m', n')$ with $m' = m/2$.

In order to solve for the new modes in Eq.~\eqref{eqboxop}, we apply a perturbative analysis for the wave equation. This involves perturbing the eigenfrequencies and eigenfunctions of the scalar field modes off of their values in the Kerr background. The modification to the eigenfrequency tells us whether the mode becomes unstable, and the evaluation for the first order perturbation in eigenfrequency should only depend on the zeroth order wave function. This fact is familiar from perturbation theory in quantum mechanics, where the leading corrections to the energy levels do not depend on the corrections to the wave function. In our case, we write the new wave function as 
\begin{align}
\label{eqscalarwf}
 \Phi_{l'm'n'}(t,r,\theta,\phi) 
 =\left [g_1(t)e^{-i\omega_R/2 t+i m' \phi}Y_{l'm'n'}(r,\theta) \right. \notag \\
 \left. +g_2(t) e^{i\omega_R/2 t-im'\phi}Y^*_{l'm'n'} \right ]e^{- \omega'_I t} ,
\end{align}
where $g_1$ and $g_2$ characterize the change in time dependence, and the perturbed wave function $Y(r, \theta)$ can be expanded as power series in $h_0$: $Y=Y^{0}+h_0 Y^{1}+...$. The unperturbed wave function $Y^{(0)}(r,\theta)$ is separable,
\begin{equation} \label{eqoriwf}
Y^{(0)}_{l'm'n'}(r,\theta) = {}_0S_{l'm'n'}(\theta)\, {}_0R_{l'm'n'}(r) \,.
\end{equation} 

We wish to solve for $g_1$ and $g_2$. The corrections to $Y(r,\theta)$ can be eliminated by defining a
 suitable, generalized inner product. This is a subtle problem here, because while the quasinormal mode 
solutions decay in time, on any fixed time slice they tend to diverge as $r$ asymptotes to infinity or 
as $r$ approaches horizon. This means that any inner product diverges if we follow a standard definition,
 keeping $r$ a real coordinate variable. Moreover, after factoring out the $t, \phi$ dependence out of the
 wave equations, we must require that $\Sigma \tilde \Box_{m,\omega}$ (the Teukolsky equation for 
scalars \cite{Teukolsky1973}) is self-adjoint with respect to this inner product. In other words, we
 require for any $\chi(r, \theta)$ and $\xi(r, \theta)$ that
\begin{equation}
\langle \chi | \Sigma \tilde \Box_{m,\omega} | \xi \rangle =\langle \Sigma \tilde \Box_{m, \omega}\, \chi | \xi \rangle\,. 
\end{equation}
 
The first problem can be solved by moving the integration contour into the complex $r$ plane. A similar integration technique has previously been used by Leaver to evaluate the amount of quasinormal mode excitation by initial data and matter sources~\cite{Leaver1986b}. The second requirement can be satisfied if we define the inner product on spin $s$ wave functions to be\\
\begin{align}
 \label{eqinnerprod}
 \langle \psi | \chi \rangle  = \int_0^{\pi}  \sin \theta d\theta \int_{\mathcal{C}} dr \,\Delta^{s} \psi \,\chi \,,
 \end{align}
where $\mathcal C$ is the complex contour for integration over $r$. In this case, the radial wave function has two branch points at $r=r_{\pm}$, and we choose the branch cuts to point vertically upward starting from the branch points, running into the upper complex plane. 
Our contour $\mathcal C$ begins in the upper complex plane to the right of the branch cut, $\Re[z] > r_+$ and a large $\Im[z]$. The contour runs down into the lower half plane parallel to the branch cut, wraps around $r_+$, and returns to large $\Im[z]$ with $\Re[z] < r_+$, running between the branch cut from $r_+$ and $r_-$ and remaining close the the former branch cut.
The asymptotic behavior of the radial functions guarantee that they decay exponentially at large $z$ in the upper half plane, which in turn guarantees that the inner product on $\mathcal C$ is finite.

The radial Teukolsky wave function is obtained analytically in two separate regions in the limit of $\epsilon \ll 1$, as discussed in e.g.~\cite{TeukolskyPress1974, Yang:2013uba}. In the inner region, where $|r-r_+| \ll M$, the approximate wave function in Boyer-Lindquist coordinates is 
\begin{align}
{}_s R_{\rm in} \propto  (- z)^{-2 i \tau/\sigma -s}\,(1-z)^{2 i \tau/\sigma-2 i \hat \omega -s }  {}_2 F_1 ( \alpha, \beta, \gamma, z)\,,
\end{align}
where $z \equiv -(r-r_+)/(r_+-r_-)$, $\sigma \equiv (r_+-r_-)/r_+$, $\tau\equiv\omega-m a/(2 r_+)$, $\hat \omega \equiv \omega r_+$, and
\begin{align}
\alpha &= -2 i \hat \omega -s +1/2+ i \delta, & \beta&= -2 i \hat \omega - s+1/2-i \delta, \notag \\\gamma &= 1-s -4 i \tau/\sigma\,.
\end{align}

On the other hand, when $|r-r_+| \gg \sqrt{\epsilon}$, the asymptotic form of the radial Teukolsky equation allows an outer solution
\begin{align}
{}_{s}R_{\rm out} =&  A\, e^{-i \omega x} x^{-1/2 - s + i \delta} \notag \\ & 
\times{}_1 F_1(1/2-s + i\delta + 2 i \omega, 1 + 2 i \delta, 2 i \omega x) \notag \\
&+ B\, (\delta \to - \delta)\,
\end{align}
where $x \equiv (r-r_+)/r_+$. Here $(\delta \to - \delta)$ means that we assign a minus sign to all the factors of $\delta$ in the previous function. The outgoing-wave boundary condition ($|r| \gg M$) forces the ratio between $A$ and $B$ to be
\begin{align}
\label{eq:Ratio}
\frac{A}{B} = e^{\pi \delta + 2 i \delta \ln(2 \omega)} \frac{\Gamma(-2 i \delta) \Gamma( 1/2 +s + i \delta - 2 i \omega)}{\Gamma(2 i \delta) \Gamma( 1/2 +s - i \delta - 2 i \omega)}\,,
\end{align}
and the overall scale of $A, B$ can be determined by comparing $R_{\rm in}$ and $R_{\rm out}$ in the matching zone: $\sqrt{\epsilon} \ll |r-r_+| \ll M$. In order to evaluate the contour integration, the above solutions are analytically continued to the complex $r$ plane, with the subtlety that there are two separate outer-solutions on each side of the $r_+$-branch cut. These two outer solutions still obey Eq.~\eqref{eq:Ratio}, but the absolute magnitudes of their $A, B$ are different from each other, according to the matching procedure. The contour integration is performed in these two outer regions and one inner region, but the result is dominated by the integration in the inner zone. Physically this means that mode-mode coupling between ZDMs mainly happens near the horizon.

We fix the overall normalization of the radial wave function in a way such that the effective mode-mode coupling strength $|H/Q|$ (with $H$ and $Q$ defined explicitly below) stays constant with varying $\epsilon$ for nearly extremal black holes. More specifically, we require that 
\begin{align}
{}_{-2}R_{lmn}(r) = |\epsilon^{-1/4-i \delta/2}|\, r^3\,e^{i \omega_{lmn} r_*},\quad r \to \infty\,,
\end{align}
where the tortoise coordinate $r_*$ is defined through $dr_*/dr = (r^2+a^2)/\Delta$, and we fix the integration constant so that to leading order in $\epsilon$, $r_* \to r + 2 \ln r$ asymptotically. 

Numerical simulations (e.g.~\cite{ZengKhanna2011}) indicate that following an inspiral the amplitude of a driving mode $h_0 \sim \mu$ at the onset of ringdown for non-extremal spins, where $\mu$ is the mass ratio of the binary. We expect this to hold in the nearly extremal case, and in our example we take $\mu = 1/8$ (a larger $\mu$ would require an accounting of backreaction). Additionally, our normalization of $R_{lmn}$ contributes a scaling $\sim \epsilon^{1/4}$ to the expected $h_0$ of a driving mode with $\delta^2 >0$. It is possible that the details of mode excitation introduce further dependence on $\epsilon$, and we can infer that this dependence does exist in the following way. 
We consider the emission from an extreme-mass-ratio-inspiral (EMRI) into a nearly-extremal host black hole. The peak emission is associated with the plunge phase, which occurs in the near-zone with some amplitude $h_{\rm max}$, which also sets the initial amplitude of the ringdown. 
In the near-zone, the ZDM wavefunctions depend on the overtone $n$, and so it is unlikely that they are collectively excited (note this goes against expectations of a power-law ringdown from \cite{Glampedakis2001,Harms2013,Yang:2013uba} who studied initial data mostly supported away from the horizon). 
This means that the individual ZDMs receive characteristic amplitudes $\sim h_{\rm max}$. 
A recent calculation of the energy flux from a near-zone orbit at fixed $z=z_0$ about an extremal Kerr indicates that the amplitude of emission is proportional to $\sqrt{z_0} \propto \epsilon^{1/4}$, and is suppressed \cite{Porfyriadis2014}. This implies a similar dependence of $h_{\rm max}$ and motivates $h_0(t=0) \sim \mu \sqrt{\epsilon}$ in our example, but more investigation is needed. We note that if $h_0$ is suppressed by larger powers of $\epsilon$, the instability may not occur.

Inserting our solution ansatz, Eq.~\eqref{eqscalarwf}, into Eq.~\eqref{eqboxop} and using our definition of the inner product Eq.~\eqref{eqinnerprod} defines equations for $g_1, g_2$ at leading order,\\
\begin{subequations}
\begin{align}
 -i m' Q \dot{g}_1
 &={g_2}H h_0(t) -m' Q \left (\omega'_R-\frac{\omega_R}{2} \right )g_1 \,, \\
im' Q^* \dot{g}_2
 &={g_1}H^*h_0(t) -m' Q^* \left (\omega'_R-\frac{\omega_R}{2} \right )g_2\, ,
\end{align}
\end{subequations}
 where 
\begin{align}
Q&\equiv\langle Y | \mathcal Q | Y \rangle\,, \qquad H \equiv \langle Y| \mathcal H( Z_{\mu\nu}) | Y^* \rangle \,, \nonumber \\
\mathcal Q &= \frac{\omega_R/2-i \omega'_I}{m'} \left [ \frac{(r^2+a^2)^2}{\Delta}-a^2\sin^2\theta\right ]-\frac{4 M  a r}{\Delta} \,.
\end{align}  
Note that $H$ has no explicit dependence on $h_0(t)$. Further, $Q$ has no explicit dependence on $m'$ to leading order in $\sqrt{\epsilon}$.
With the ansatz $g_j=A_j e^{\int^t \alpha(t') dt'}$ ($j=1,2$) (with $A_j$ to be determined) and the requirement
of obtaining a  non-trivial solution to the above system, one obtains Eq.~\eqref{eq:alpha} for $\alpha$ and the condition~\eqref{eqcri} for mode growth.

\bibliography{scibib}

\bibliographystyle{apsrev}

\end{document}